# Flexibility potentials of a combined use of heat storages and batteries in PV-CHP hybrid systems

Tanja M. Kneiske [a]1, Martin Braun[a,b]

[a]*Fraunhofer Institut for Windenergy and Energysystemtechnology (IWES), Königstor 59*
*34119 Kassel, Germany*
[b]*University of Kassel, Department of Energy Management and Power System Operation, Wilhelmshöher Allee 73, 34121 Kassel, German*

**Abstract**

**Due to the 2012 change in the renewable energy act the feed-in tariffs and as result the number of newly installed photovoltaic systems decreased dramatically in Germany. Therefore, there was the need, particularly, in the residential sector to develop new business ideas for photovoltaic systems. In context of this development, combined photovoltaic and heat-power systems were analyzed, which provide not only electricity but also heat throughout one entire year. Flexibilities are provided in form of thermal and electrical storage systems leading to many possible setting options requiring an elaborated control management. In this paper, a new optimized control algorithm is proposed that in contrast to standard strategies can operate optimal even under incorrect weather and load forecasts. A predictive controller based on a mixed integer optimization problem and an additional so called secondary, rule-based controller is combined. The controller is based on several input data. Depending on the choice of these data the PV-CHP hybrid system can be used for different control strategies, like maximum self-consumption, minimum $CO_2$ emission or minimum operational costs. The main focus in this paper is to study if the flexibility potentials of the combined thermal and electrical storage systems can ensure a market oriented or a grid-friendly behavior. This is studied in five control options. As a result we found that the control algorithm is stable and able to adapt to the different conditions. In conclusion, it was discovered that the storage systems play a crucial role in terms of forecast differences and parameter changes. Storage systems are as expected the key-element for the flexibility of PV-CHP hybrid systems.**

## 1. Nomenclature

| | |
|---|---|
| PV | radius of |
| CHP | position of |
| VPP | further nomenclature continues down the page inside the text box |
| PCC | Point of common coupling |
| TES | Thermal energy storage |
| SOC | state of charge |
| EEG | Renewable Energy Act, Erneuerbare-Energien-Gesetz |
| KWKG | Heat-Power Cogeneration Act, Kraft-Wärme-Kopplungs-Gesetz |
| KFW | Germany's development bank, the Kreditanstalt für Wiederaufbau |
| MPC | Model predictive control |
| th | thermal |
| el | electrical |

## 2. Introduction

The energy transition in the German energy grid has led during the last 5 to 10 years to substantial changes in the generation structure from large power plants to decentralized small generators. A key component among decentralized small generators plays renewable energy sources. According to the German Federal Ministry for Economic Affairs and Energy these renewable energy sources accounted for more than 31 % of the gross electricity consumption in Germany in 2015.

* Corresponding author. Tel.: +49 561 7294 136; fax:+49 561 7294 200.
 *E-mail address:* tanja.kneiske@iwes.fraunhofer.de

Among these decentralized generators more than 1.5 million are installed small distributed photovoltaic (PV) systems. Many of these are roof systems on private homes and commercial buildings. The feed-in tariff dropped to 12.31 Eurocent/kWh. As a result, new business cases, like PV-battery systems, emerge which promise an economic benefit due to a higher self-consumption and/or a grid-friendly behavior [1], [2], [3].

While PV systems are mostly generating energy during the summer period, standard combined heat and power (CHP) systems are able to produced electrical and thermal energy throughout the year, but mostly in winter. In 2015 11 % of the gross electricity consumption in Germany was due to general CHP-systems [4]. It has been shown that CHP system with an intelligent controller can lead to a better grid integration, provide grid services or can be part of a virtual power plants (VPP) [5], [6], [7].

The combination of both systems seems to be a promising step towards a combined electrical and thermal power supply. An increased use of the storages systems, in particular the use of the battery not only in summer for the PV energy but also during the other seasons for the CHP plant could improve the utilization of the battery. The authors in [8] have shown by analyzing five locations in Spain that the primary energy consumption can be reduced when using a hybrid system. However, when it comes to live cycle costs they also found out that a conventional system (electrical grid plus natural gas boiler) is still superior. Another live cycle analysis has been done in [9]. Compared to a conventional electricity supply from the grid and heat from a natural gas boiler an overall significant improvements in all environmental impacts compared to the conventional energy supply, ranging from 35 % for depletion of fossil fuels to 100 % for terrestrial ecotoxicity has been found. In [10] the authors studied the sizing of the components of a hybrid system. They found that in contrast to solar technologies, the size of micro-CHP units is heavily influenced by several factors and parameters, such as the investment costs, energy loads and tariffs. Outcomes suggest that a hybrid system can reduce the primary energy consumption of households more than single PV technology, but its size must be properly identified for existing costs and conditions. In [11] the authors studied PV-CHP hybrid systems in 30 households to derive the self-sufficiency proportion, grid demand profiles and economic costs for the consumers. They found that such a system is on the one hand increasing the self-sufficiency proportion and reduces critical parts of grid demand profiles in terms of ramp-ups. On the other hand, however, they could also show that these systems are only beneficial supplying high energy demands (> 4300 kWh/yr).

In the following the flexibilities of a PV-CHP hybrid system in respect of a market and grid-friendly behaviour is investigated. This is done by studying the power at the point of common coupling (PCC). A minimal cost strategy for the PV-CHP hybrid system is used. Contrary to the central control of a VPP, the control strategy and therefore the impact on the grid can be influenced by incentives and curtailment. Constant and time-dependent electricity prices and feed-in tariffs are studied in different options. In order to meet these requirements, a two level control algorithm was developed. The primary control is based on a model predictive control (MPC) optimizer which calculates cost minimized time schedules for the next six hours with a 10 min temporal resolution. Within these 10 min a rule based control is used to account for the differences between the forecasted set values and the actual values.

The paper is structured as described hereinafter. The methodology with its used assumptions and parameters are described in section 2 introducing two modes and five options for further discussion. The simulation results, like daily profiles and total operational costs are outlined and assessed in Section 4. A conclusion of the calculated results is outlined in section 5.

**3. PV-CHP hybrid system**

This section gives a description of the components and the combined control algorithm. Afterwards, two modes are defined to show the effect of using a forecast method. In the end of this section we motivate the choice of five incentive options which will be discussed in the section thereafter.

*3.1. Components*

The components considered in the following study are shown in Figure 1. The system includes thermal and electrical components for production devices, storages, loads and grid exchange.
The assumed parameters are shown in Table 1.

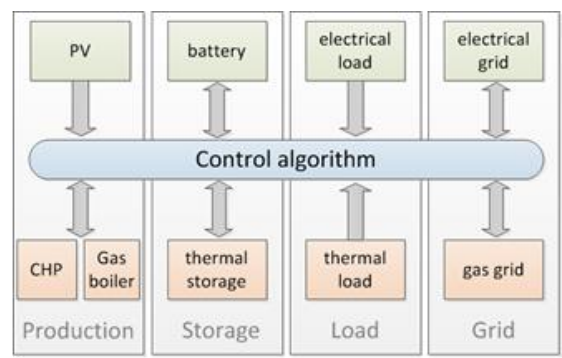

**Figure 1** The control algorithm includes the shown electrical and thermal components. The controller has no influence on the PV, electrical and thermal load components.

Table 1 Parameters for the components of the PV-CHP hybrid system.

| System | Size / max. Power | Efficiency |
| --- | --- | --- |
| PV | 3.2 kWp | 1 |
| CHP | 1.0 kWel (on/off) | 0.263 (el), |
|  | 2.4 kW$_{th}$ | 0.657 (th) |
| Battery | 4 kWh (60% usable) | 0.99, 0.9, 0.92 |
| Thermal energy storage (TES) | 300 l | 0.98, 0.9, 0.92[1] |
| Natural gas boiler | 2.4 – 30 kWth | 1 |

PV profiles are derived from measured data in the Kassel region in 2013. The load profile has been calculated using the time series of guidelines of the Association of German Electro-engineers VDI4655[2]. The VDI4655 dataset contains electrical and thermal household loads for single family houses and 12 prototypical days. They can be combined to rebuild the energy demand profiles for the entire year 2013 with a ten minute resolution for the region Kassel to be consistent with the PV profiles.

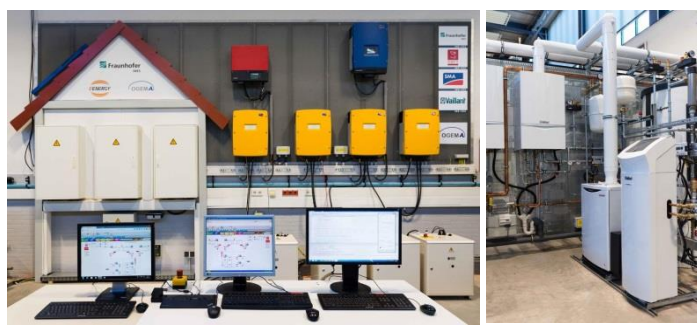

Figure 2 "Ine-Ves" test environment of a PV-CHP hybrid system. Left panel: Converters and the batteries are shown in the back. One of the computers is running the OGEMA energy management system, containing the optimizer, while the other computer holds the control unit for the entire habrid system tests. Right panel: the gas boiler and the Vaillant ecoPOWER 1.0 are shown. The thermal storages and the pipe system representing the different heat cycles are behind the metal wall construction.

In the following study only the operational costs are taken into account, because the focus in this work is to develop a control strategy. The size of the components is given by the available products provided by Vaillant GmbH, SMA Solar Technologie AG and Saft Batterien GmbH. The PV-CHP hybrid system being object of this study is built up at the Fraunhofer IWES DeMoTec laboratory in Kassel, Germany. The control algorithm shown here is implemented on an energy management system, and is tested right now in a "hardware in the loop" environment (see Figure 2).
The combined controller is shown in Figure 3.

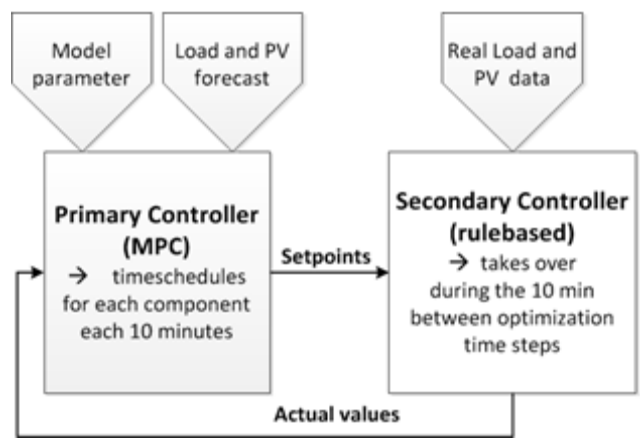

Figure 2 Schematic depiction of the combined control algorithm consisting of a primary controller based on a model predictive control and a secondary rule-based controller

The primary controller is a MPC which is based on an optimizer solving a mixed integer linear problem (similar to [12]). The default setting is a prediction horizon of 6 hours. Other timespans between 2 and 24 h were also tested, rendering similar results. Longer timespans, however, resulted in increased computation time and seem therefore insufficient. A review of optimized control strategies for energy systems in buildings is summarized by [13].
The secondary controller is based on constraints. The constraints are defined to account for differences between forecast set values and real values. The secondary control algorithm in the simulation is based on four steps which are orientated on the already build-in controller of each single system component. The build-in controller

---
[2] www.vdi.de

(e.g. battery management system, BMS) of the components should remain running, while the new combined control is put on top by an energy management system. By this way already existing components can be used for a PV-CHP hybrid system keeping the development costs low. The steps are shown in more detail in Figure 4.

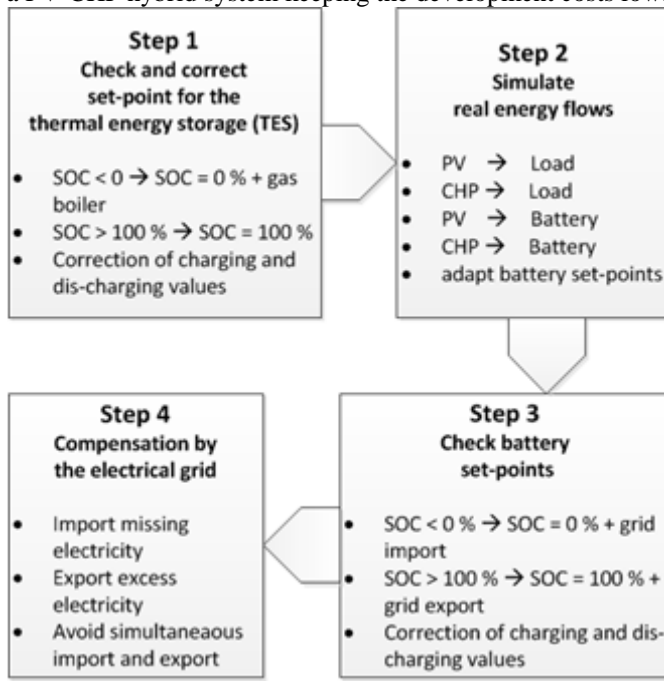

Figure 3 Four steps of the secondary rule-based controller, which is used between the optimization time steps.

In the first step of the secondary control the set values for the thermal storage are checked and corrected. The CHP system and the gas boiler are controlled by the temperature level of the thermal storage. Afterwards, the real energy flows are simulated, from the real PV output to the real electrical load and from the CHP system to the real electrical and thermal loads. This can change the charge and discharge set values of the battery which need to be corrected as well. Finally, the missing and excess electrical energy is equalized using the electrical grid. The SOC of the thermal and electrical storage are corrected based on the charge and discharge values. The running mode of the CHP is not changed. If the CHP is on, it stays on, when it is off it stays off. The CHP can only be turned on and off by the primary controller. If the thermal demand is much smaller than expected by the forecast and because the CHP cannot be turned off by the secondary controller the TES SOC could be larger than 100 %. For further details see [14].

In the following we will compare two modes to understand the effect of the chosen forecast method on the PV-CHP hybrid system:
- Mode A: Perfect Forecast
- Mode B: Persistence Forecast

*3.1.1. Mode A - Perfect Forecast*
The first mode considered is the default mode, where we assume that the PV electricity production and the load demand are exactly predictable. This mode will never occur in a real hybrid system, but it is used to test the primary controller without the correction by the secondary controller and the effect of the forecast method. This mode is mostly analysed in literature by other authors so far.

*3.1.2. Mode B - Persistence Forecast*
Mode B allows studying the effect of forecast uncertainties on the time schedules, costs and other parameter of the PV-CHP hybrid system. We assume that forecast and real input values differ, which is the normal case for an installed hybrid system in a real environment. The forecast method used is a persistence forecast, assuming the values today are going to be the same as yesterday at the corresponding time. In general, it can be observed that this forecast method leads to a good approximation of the total energy required throughout one day. Although the peak power is predicted very well the forecast fails to set the exact time of occurrence of the peaks.

A running mean forecast method using the sliding average of the last three days has been tested as well. The results were compared to the persistence forecast. We found out that the running mean forecast is less useful, since it is reducing the peaks. This is eventually resulting in lower total energy demand and imprecise running times of the CHP plant. Therefore, in the following only the two Modes A and B are investigated.

## 4. Incentive options

The PV-CHP hybrid system and the combined controller introduced in the last section can be analysed for different control strategies, e.g. maximum self-consumption, minimum $CO_2$ emission [15] or minimum operational costs. In the following only the last strategy of minimum operational costs is considered. This strategy is very flexible due to the possibility of using incentives to influence the PV-CHP hybrid system towards a market-oriented or a grid-friendly behaviour, which is the focus of this work.

For this purpose, five different options are distinguished. The first two options are defined to depict the most common situations in Germany today. The first option is based on the current laws including feed-in tariffs and the second option includes no incentives at all.

- Option 1 – EEG/KWKG-oriented incentive
- Option 2 – no incentives

The next three options are chosen to study the capability of a PV-CHP hybrid system for PV-integration on a summer-day and its impact on the grid.

- Option 3 – curtailment (50 %)
- Option 4 – variable electricity price
- Option 5 – variable feed-in tariff

The options are described in more detail in the following.

### 4.1.1. Option 1 – EEG/KWKG-oriented incentive

The input variables are motivated by the Renewable Energy Act (EEG, Erneuerbare-Energien-Gesetz) and the Heat-Power Cogeneration Act (KWKG, Kraft-Wärme-Kopplungs-Gesetz). The following feed-in tariffs and costs have been assumed accordingly:

**Table 2 Assumed costs and sales.**

| Costs | Euro/kWh | Sales | Euro/kWh |
|---|---|---|---|
| gas costs | 0.0652 | PV feed-in | 0.1256 |
| electricity costs | 0.2838 | CHP feed-in | 0.09392 |
| costs for CHP cold start | 0.02 | Avoided grid costs | 0.005 |

### 4.1.2. Option 2 – no incentives

Feed-in tariffs are covered by EEG levy which lead to higher electricity costs. However, already first changes have been passed, reducing these and instead, moving back to market control mechanism. Fixed feed-in tariffs are therefore only a temporal situation. Conclusively, a long-term requirement is a PV-CHP hybrid system which is economic without any incentives set by an EEG or a KWKG. In this option the possibility is studied cutting out all benefits, leaving a system with the costs from table 2 only.

### 4.1.3. Option 3 – 50 % Curtailment

Owners of PV-battery system get a KfW (Germany's development bank, the Kreditanstalt für Wiederaufbau) subsidy if they accept a curtailment of 50 % at the point of common coupling. Motivated by the possibility to get the subsidy for the PV-CHP hybrid system the 50 % value will be used to test and analyze its impact on the control strategy.

### 4.1.4. Option 4 – variable electricity price

Fixed electricity prices are still the common situation of private households. A change from fixed to variable electricity prices and the possibilities for PV systems were addressed in publications [1], [16] investigating if variable prices, which are based on energy market fluctuations, are able to reduce electricity costs for end-consumers. In context of this issue, variable prices from EEX Spot Market in 2013 are used to study if the PV-CHP hybrid system is able to follow the changing input parameters. The price curve is depicted in Figure 9 together with the results.

### 4.1.5. Option 5 – variable feed-in tariff

In option 3 on the one hand a fixed curtailment is studied which leads to a peak shaving of PV-power. On the other hand, option 2 assumes fixed feed-in tariffs. In this last option, both ideas are tried to combine, by introducing variable feed-in tariffs. The aim is to realize a smaller PV-peak at noon. Therefore, the feed-in tariff is set to zero between 10 and 2 o'clock during noon. Note that only the PV feed-in tariff is considered variable. The feed-in tariff for the CHP plant is not changed. All these options are just examples. Other incentives can be tested in the given simulation framework.

## 5. Result and discussion

In this section, the results for the defined options under the different predefined conditions are shown. First, the results of the component behavior for the different options during one day are illustrated as daily profiles. Afterwards, the costs of the different options are compared and discussed.

*5.1. Option 1 – EEG oriented incentive*

The option with an assumption of feed-in tariffs for PV and CHP systems is leading to a control strategy where a high amount of the produced energy is fed into the grid. In summer the exported energy is dominated by the PV generation, in winter the energy of the CHP system is exported. Important to notice is that in the transition-period the energy of both systems (see Figure 5) is fed into the grid leading to higher grid integration problems than PV and CHP alone. The electrical load is provided by the CHP plant (Figure 5, upper panel, dark green), while the peak loads are covered additionally by the battery. The results for Mode A and Mode B are shown in Table 3. The uncertainty of forecasts in Mode B leads to higher costs about 5% and to higher number of battery cycles. As another reference value, a conventional system (CS) consisting of solely the electrical grid and a gas heater is used. For electrical and thermal energy demand of 11.4 kWh and 85 kWh respectively the total energy costs sum up to 8.76 Euro, with 5.52 Euro for heating and hot water and 3.24 Euro for electricity. Note that costs for additional losses due to efficiencies in the system have not been taken into account for the conventional system.

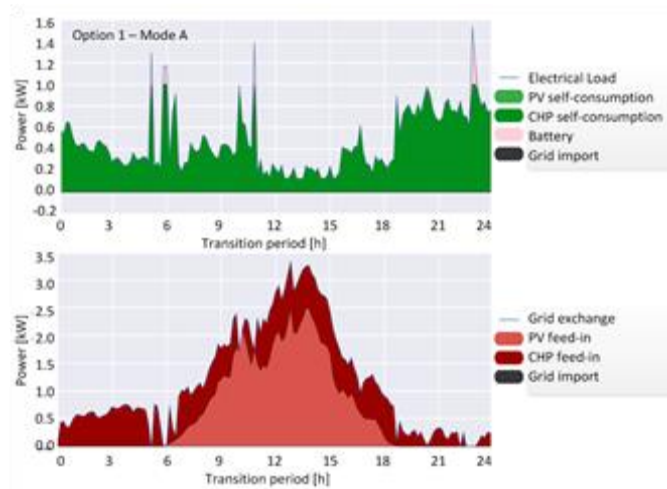

**Figure 4 Upper plot: electrical load profile with a share of components for 24 h. Lower plot: grid exchange at the point of common coupling. Positive values define energy export, negative values energy import.**

*5.2. Option 2 – No incentives*

The results of this option show the typical behaviour of a system with a maximum self-consumption rate as there are no benefits of exporting energy. The electrical load demand is supplied by the PV-system during day time, while at night the CHP system provides the needed energy (see Figure 6, upper panel). In the early morning and late evening hours the charged battery is discharged. Only the excess energy of the PV system is fed into the grid. Compared to option 1 the exported peak power of the PV-CHP hybrid system is reduced from almost 3.5 kW to values below 2.5 kW which would lead to a more grid-friendly behaviour for the chosen day (see Figure 6, lower panel).

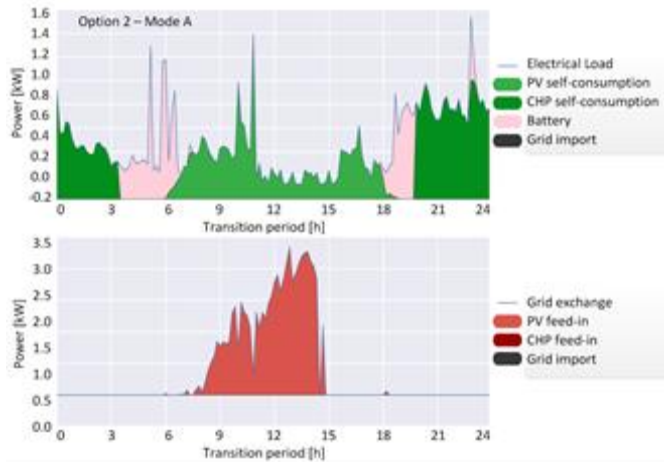

**Figure 5** Upper plot: electrical load profile with a share of components for 24 h. Lower plot: grid exchange at the time of day of common coupling. Positive values indicate energy export, negative values imported energy.

At the end of the sunshine period the PV energy is used to charge the battery (around at 18 h). The CHP system is turned on around 20 h when the battery is empty until around 3 h in the morning (dark green) when the battery is full again. The charge of the battery is sufficient enough to fill the time gap until the PV-system is providing enough energy again. In summary, if no incentives are being paid for exported energy, the controller switches to a mainly pure self-consumption strategy by using the flexibility possibilities of a battery. The results in Table 3 show that the self-consumption rate of the CHP is at 100 %, while the PV self-consumption rate is around 40 %, which is comparable to a PV-Battery system. The running time of the CHP plant is reduced from 24 h in option 1 to 8 hours in option 2, including the forecast uncertainties in Mode B even to 4 hours for the day. The battery is used more often from less than 1 to 1.55 cycles a day in option 2 and mode B.

*5.3. Operational costs*

The operational costs are calculated for 24 h and shown in Figure 7. Since the control strategy depends strongly on the investigated season, thus the outside temperature and solar radiation, the results presented here are evaluated for three different days, a typical "winter", "summer" and "transition period" day. The day types are chosen to test the optimizer under different exogenous situations. The "winter-day" is characterized as follows: Besides electrical and hot water demand, also space heating demand due to cold outside temperatures is high. The PV-energy-production is very small, nearly zero. The "summer day" in comparison shows no space heating demand but a high PV energy production output. The "transition period day" reveals the coexistence of space heating demand and PV energy production output. The total operational costs of both options are smaller compared with the total costs of a conventional system (see Figure 7).

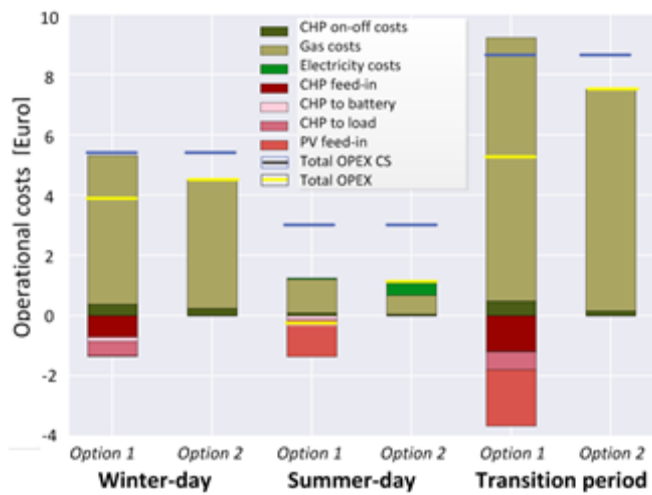

**Figure 6** Total operational costs (yellow lines) with a share of components for 24 h. Option 1 and 2 are compared for three different days, representing three different seasons. Positive values depict costs, while negative values represent benefits. The values for a CS are calculated for the conventional system and shown as blue lines.

Option 2, without incentives leads to higher operational costs compared to option 1 for the same needed electrical and thermal energy. Furthermore, Table 4 also reveals the seasonal impact: based on the information given in table 2 in case of option 1 compared to the "summer-day" the costs increase by 14 % on the "winter-day" and by 25 % on the "transition period day", respectively. In case of option 2 during winter time the income of the CHP feed-in is lost and cannot be completely compensated by the subsequent reduced gas costs. On the "summer-day" the 12 Eurocent profit is lost comparing option 1 and 2 and the needed energy is covered by the gas and the electrical grid leading to costs of around 1 Euro. The "transition period day" is the most unfavorable time, since here both feed-in benefits are lost comparing option 1 with 2.

Comparing the costs of the conventional system with the values computed here we find that for both options the operational costs are 36 % smaller for option 1 and 14 % smaller for option 2.

*5.4. Option 3 – 50 % Curtailment*

If an additional rule is added to the optimizer which forbids to feed-in more than 50% of the installed peak power of the PV-system, the battery is used to store the excess energy. The proposed combined control algorithm is capable of including curtailment (see Figure 8). The stored energy is used later in the afternoon to cover the electrical load. The energy is not lost, but is automatically charged to the battery to prevent grid consumption later.

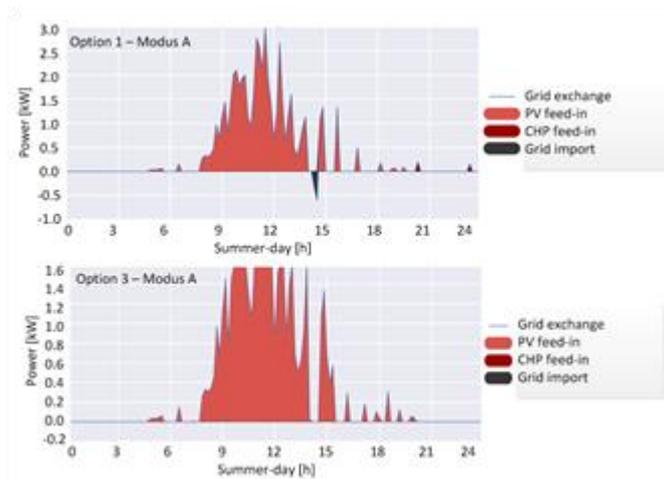

**Figure 7 Grid exchange at the time of day of common coupling. Positive values indicate energy export (red), negative values imported energy (black). The export is dominated by PV power. In the lower plot the curtailment of 50 % leads to a cut-off at 1.92 kW.**

*5.5. Option 4 – Variable electricity prices*

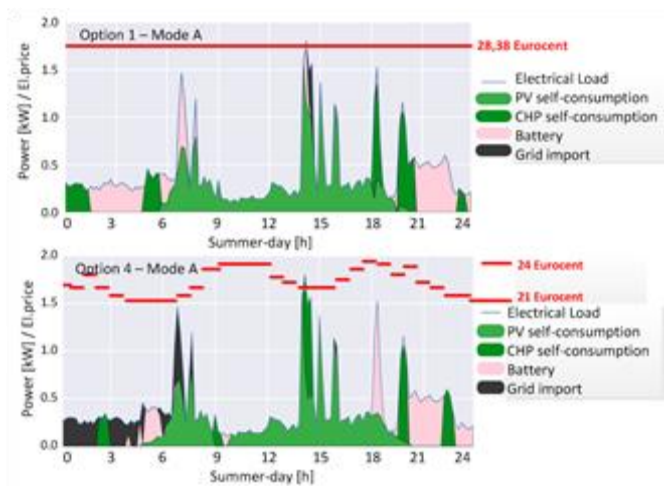

**Figure 8 Electrical load profile with component share for 24 hours. Upper plot: with constant electricity price; lower plot: with variable electricity prices.**

Option 4 considers the impact of variable electricity prices. As seen in Figure 9, by taking into account variable electricity prices a higher amount of the energy is imported from the grid.  During the first hours of the day until 6 h the consumption is mainly fed by the grid due to cheap electricity prices. (see Figure 9). The rising price until noon and in the evening around 18 h is resulting in a self-consumption strategy Note that in this option the average energy price is in general lower than in option 1, representing energy market standards in 2013. This results in a larger sensitivity of the controller to variable prices. The amount of PV energy production fed into the grid has not changed compared to Option 1. The same total amount of PV energy including all peaks up to 3 kW is exported to the grid. Therefore, no grid-friendly behavior can be seen using this scenario.

In summary, it can be observed, that a minimum value of     21 Eurocent up to a maximum value around 24 Eurocent is sufficient to change the control strategy of the PV-CHP hybrid system, but does not lead to the desired change of a reduced power-flow at the point of common coupling.

*5.6. Option 5 – Variable feed-in tariff*

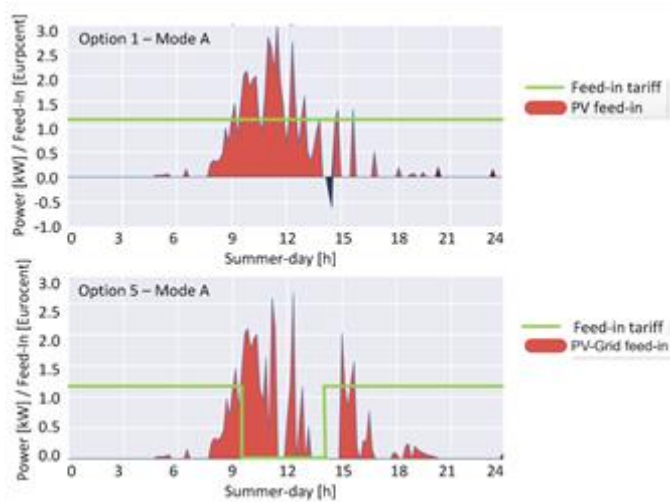

**Figure 9 Grid exchange at the time of day of common coupling. Positive values show the energy export, negative values imported energy. Upper panel: with constant feed-in tariff, lower panel: with variable feed-in tariff.**

Variable electricity prices have no effect on the PV-peak. Therefore, contrary to that, the option with a variable feed-in tariff is expected to change the exported amount of energy for the chosen summer day. In this option a variable PV feed-in tariff is tested, while the CHP feed-in tariff is kept constant. In Figure 10 the results of this option reveal, that a simple approach of no feed-in benefit from 10 h to 14 h is already providing a peak reduction of about 80 % of the installed PV capacity for the chosen summer day. On a day of the transition period a variable CHP feed-in tariff could reduce the grid load even more since PV and CHP feed-in tariffs can be controlled separately. However, this is out of scope of this study and part of further research. Further studies could also include optimized feed-in profiles for all seasons.

*5.7. Operational costs*

The operational costs on a summer day for all five options are shown in Figure 11. The largest share is the gas costs. Gas is consumed by the CHP system and the gas heater. CHP cold starts play only a minor role. Two options show a significant amount of electricity costs. This is in option 2, without incentives, and option 4, using variable electricity prices. When there is a feed-in tariff > 0 large amounts of PV-energy are exported to the grid resulting in benefits of around 1 to 1.4 Euros. The other profits play a minor role. Looking at the operational costs and benefits of the chosen day for the different options with incentives, the total amount is near zero. As a result, we see that the total costs in summer are not varying much when comparing the different options. Thus, even when scenarios like curtailment, variable electricity prices or variable feed-in costs are introduced, the results seem to be stable. The flexibilities due to three generators and two storage systems are large enough to cope for the effects caused by these different scenarios. Only the option without any incentives leads to higher costs of around 1 Euro per day in summer.

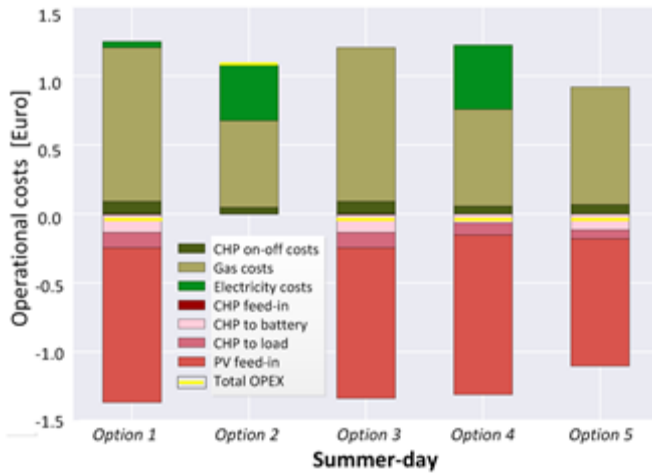

**Figure 10 Total operational costs with the components share for 24 h. Option 1 to 5 are compared for a summer day. Positive values depict costs, while negative values represent benefits. The total costs are shown as yellow line. The total costs of 3 Euro for the conventional systems are much higher than the section plotted here.**

**6. Conclusion**

The PV-CHP hybrid system with the new combined control algorithm is able to adapt to different options of incentives. The combination of the several components, in particular the storage systems, can improve the flexibility potentials in all seasons.

In the Option 1, Mode A with a perfect forecast, the battery is not used very often, with 0.1 full battery cycles for the chosen day. But as soon as forecast derivations are included (Mode B), the battery is used more often, indicating a need for flexibility. In general, the more restraining conditions are proposed to the system, the higher the number of battery cycles. Numbers range from 0.5 to almost 3 cycles per day, depending on the option, the mode and the season.

Incorrect forecast values are compensated by the battery and the thermal storage but lead to 5 % to 10 % higher operational costs.

The study of grid friendly control strategies of a PV-CHP hybrid system in summer shows that option 5, with variable feed-in tariffs, has promising results and should be studied together with variable CHP feed-in tariffs in the future. To reduce the exported power also a curtailment of 50 % is possible without higher operational costs, because the battery can store the excess energy for later consumption.

The control algorithm presented here has been implemented on an energy management system controlling real PV CHP hybrid system in the DeMoTec laboratory in Kassel.

**Acknowledgements**

The authors thank the German Federal Ministry for the (BMWi) and the Forschungszentrum Jülich GmbH (PTJ) for the support within the framework of the project "INE-VES" FKZ 0325561A.